\documentstyle[12pt]{article}
\title{Study of the band--gap shift in CdS films: Influence of thermal
annealing in different atmospheres.}

\author{S.A. Tom\'as, O. Vigil$^*$, J.J. Alvarado-Gil,
R. Lozada-Morales,\\[-.1em] O. Zelaya-Angel\\[-.1em]
Departamento de F\'{\i}sica\\[-.1em]
H. Vargas$^{\dag}$ \\[-.1em]
Programa Multidisciplinario en Ciencias Aplicadas y \\[-.1em]
Tecnolog\'{\i}a Avanzada.\\[-.1em]
Centro de Investigaci\'on y de Estudios Avanzados del IPN\\[-.1em]
Apdo. Postal 14-740, M\'exico D.F. 07000, M\'exico\\[-.1em]
A. Ferreira da Silva\\[-.1em]
Instituto Nacional de Pesquisas Espaciais-INPE\\[-.1em]
Laboratorio Associado de Sensores e Materiais-LAS\\[-.1em]
CP 515,12201-970 Sao Jose dos Campos, S.P., Brazil}

\date{  }
\begin{document}
\maketitle
\vspace*{-1cm}
\begin{abstract}
We study by photoacoustic spectroscopy the band--gap shift effect of CdS
films. The CdS films were grown by chemical bath deposition and exposed to
different annealing atmospheres over a range of temperature in which the
sample structure  is observed to change. We show the band--gap evolution as a
function of temperature of thermal annealing and determine the process which
produces the best combination of high band--gap energy and low resistivity. It
allows us to know a possible procedure to obtain low--resistivity CdS/CdTe
solar cells with high--quantum efficiency.
\end{abstract}

\noindent
$^*$ Permanent address: Physics Faculty, University of La Havana, La Havana,
Cuba.\\
$^{\dag}$ On leave of absence from the Universidade Estadual de
Campinas, S.P., Brazil.\\[1em]
PACS N. 43.85, 62.65, 78.20.D, 78.65, and 81.40.E
\newpage
\baselineskip 24.1pt
\noindent {\Large Introduction}\\

The study of heterojunctions and their use in a large variety of technological
applications has become one of the principal subjects in solid state research.
Among other applications, we mention the use of heterojunctions in transistors
and diodes, as well as in other optoelectronic devices. Likewise, it is well
known the importance represented by heterojunctions in solar cells
development. In particular, CdS/CdTe heterojunctions, represent one of the
principal candidates for low cost and high conversion-efficiency solar cells
\cite{Ferekides}. In this type of solar cells, the obtention of CdS films with
low resistivity plays an important role because it helps to diminish the
dipositive sheet resistance and obtain the space charge region in the active
zone, i.e., in the CdTe film.\\

It is also known that CdS samples present two different structural phases,
namely, the highly stable hexagonal phase and the metastable cubic phase.
Even though it is possible to know by means of reflectivity measurements at
room--temperature that these two phases have a band--gap energy differing less
than 0.1eV, it is not possible, however, to infer any other conclusion about
them \cite{Cardona}.\\

CdS films present a band--gap shift (BGS) effect when exposed to certain
experimental conditions. This effect was reported by Balkanski {\em et al.}
\cite{Balkanski} for Si films as a function of temperature. Later on, because
of its technological importance in optoelectronic design, many researchers have
investigated the BGS of intrinsic and extrinsic semiconductors
\cite{Berggren}-\cite{Wagner}. Recently, Zelaya--Angel {\em et al.}
\cite{Zelaya} have reported the first investigation of BGS as a function of
temperature of thermal annealing (TTA) for CdS films. Nevertheless, it is
worth saying that such a BGS effect, as well as the growth procedures to
obtain CdS films in either of the phases, are not well understood up to now.\\

In this work we investigate by photoacoustic spectroscopy (PAS) the band--gap
shift effect of CdS films as a function  of
TTA in different annealing atmospheres. By extension, we show the interesting
and technologically important combination of low resistivity CdS/CdTe solar
cells with high energy-gap. The annealing atmospheres investigated are Ar,
Ar+S$_2$, H$_2$~+~In, H$_2$, and air, each of them fixed at certain
temperatures within a range from 200$^o$C to 450$^o$C. The purpose of studying
it by PAS is that this technique allows us to obtain spectra which clearly
show the BGS effect \cite{Zelaya}-\cite{Vargas}.\\

\noindent{\LARGE Experimental.}\\

The approximately 0.25$\mu$m thickness cadmium sulfide films were deposited by
the chemical bath deposition method (CBD). This deposition occurs due to an
ionic precipitation of salts dissolved in an aqueous solution. The salts used
were CdCl$_2$, K0H, NH$_4$NO$_3$, and CS[NH$_2$]$_2$ (thiourea). The
solution was kept at (80$\pm$3)$^o$C and continuously stirred to guarantee an
homogeneous distribution of the chemical compounds, among other effects
\cite{Kaur}.\\

The postdeposition  thermal treatments given to the samples were performed
in a hot wall furnace with a quartz tube. Taking advantage of the furnace
temperature profile, CdS samples were simultaneously placed within a range of
temperature from 200$^o$C to 450$^o$C. For Ar gas and Ar gas + S$_2$ vapor
flux annealings, CdS samples were placed at 208, 260, 353, 398, and 447$^o$C
(419$^o$C the last for Ar+S$_2$) during 28h. The S$_2$ vapor flux was obtained
by placing a boat containing sulfur at 190$^o$C. For H$_2$ gas, H$_2$ gas + In
vapor flux, and air annealings, CdS samples were placed at 200, 250, 300, 350,
400, and 450$^o$C during 1h. In the H$_2$+In case, a boat containing In was
placed at 630$^o$C, whereas for the H$_2$ annealing, an H$_2$ flow was kept
for 30 min. before placing the samples. The stimated pressure for all the
treatments was about 1 atm. \\

The photoacoustic absorption spectra of the films were obtained in the region
400--700nm by use of a standard photoacoustic spectrometer. This spectrometer
is
fitted with a 1000W Xenon lamp (Oriel), whose radiation is focused onto a
variable--frequency light chopper set at 17 Hz. The photoacoustic signal is
preamplified before providing the input to the signal channel lock--in
amplifier.
The resultant photoacoustic spectra are recorded in a computer, which
simultaneously display the wavelength--dependent signal intensity. \\

\noindent{\LARGE Results and discussion.}\\

In Fig.1 we illustrate the  PA spectra corresponding to both the as--grown CdS
sample and the whole set of CdS films annealed in Ar+S$_2$ atmosphere. We
assume that the value of the band--gap energy is given by the energy which
matches the spectrum  inflexion point. With this assumption, we observe the
BGS effect as a function of TTA. The values of the band--gap energy (BGE)
for the samples annealed in Ar and Ar+S$_2$ atmosphere are shown in Table I,
whereas the ones for H$_2$, H$_2$+In, and air are shown in Table II.
Focusing us on the Ar+S$_2$ annealing we obtain the BGE, as shown in Fig.2.
We observe that the BGE starts decreasing its value as a
consequence of the thermal annealing. This effect carries down the BGE from
its 2.42 eV initial value at 80$^o$C to 2.28 eV at roughly 350$^o$C, where a
minimum is reached. Above this temperature the BGE presents an opposite
behavior, increasing its value until 2.35eV at 418$^o$C. This
narrowing--widening--like behavior presented by CBD/CdS films annealed in
Ar+S$_2$ atmosphere has been interpreted as a cubic to hexagonal phase
transition, which occurs at the  temperature corresponding to the minimum BGE
\cite{Zelaya}. The TTA coefficients of the BGE have also been approximately
found in the linear part of the data \cite{Burger} with values of about
$\mp 10^{-4}$ eV/TTA($^oK$), corresponding respectively to the narrowing and
widening region. The magnitude of this coefficient agrees generally with those
of other reported materials, e.g., Si, Ge, GaAs, and MgI$_2$
\cite{Burger},\cite{Sze}.\\

With regard to the other annealing atmospheres, we show in Tables I and II
that all of them start with the same behavior as the Ar+S$_2$ case, that is,
the samples present a narrowing--like behavior. Some of these annealings
decrease the BGS faster of slower and reach a higher or lower minimum value
than Ar+S$_2$. Nevertheless, it is very important to point out that none
presents a clear widening--like behavior as does Ar+S$_2$. On the contrary,
all of them show a stable tendency above the temperature corresponding to the
minimum BGE. These results are illustrated in Fig.3 where the BGS, given by

\begin{equation}
\Delta Eg(T) = Eg(80^oC) - Eg(T),
\end{equation}
\noindent
is plotted versus the TTA. Here, $Eg(80^o$C) is the band--gap energy of the
as--grown samples. We notice that, among the investigated annealings, the
highest $\Delta Eg$ values are produced by the air, the maximum $\Delta Eg$ =
0.17 eV taking place at roughly 350$^o$C. Moreover, only the Ar+S$_2$ annealing
produces a considerable diminution of the BGE after having reached its
maximum value.\\

Owing to our interest in finding the thermal process which
produces CBD/CdS films with the best combination of high band-gap energy and
low resistivity, we deal now with the sample electrical resistivity as a
function of TTA. As all the annealing atmospheres produce similar
resistivity effects in the samples, we take advantage of it and describe
qualitatively their general behavior.
Electrical measurements were performed by the two probes method for high
resistivity films and the four probes method for low resistivity films at room
temperature. The four probes technique was not satisfactory for high
resistivity films because of the same order of magnitude of the film resistance
and the electrometer input impedance. Resistivity of the as--grown samples
was measured to be $\rho = 2.3\times 10^7 \Omega$ cm.  Subsequently, as
the annealings were performed, the resistivity decreased with increasing TTA
until reaching a minimum value, then increasing again. As can be seen in
Table III, $\rho(T)$ takes different minimum values for each different
annealing atmosphere.\\

The different minimum values of $\rho$ are analysed below. The Ar+S$_2$
annealing diminishes the resistivity by increasing the grain size, therefore
decreasing the number of grain boundaries in the films. On the other hand, the
Ar annealing, besides of in
creasing the grain size, originates a high amount of sulfur vacancies as a
consequence of non-equilibrium conditions. In the air annealing, the presence
of oxygen creates both CdO and SO$_4$Cd layers on the surface, increasing
slightly the resistivity of
the films respect to the Ar
case \cite{Kolhe}. The thermal treatments in H$_2$ and H$_2$+In present the
advantage of containing H$_2$, which is a strong agent for grain boundary
passivation by oxygen chemiabsorption \cite{Melo},\cite{Chopra}, decreasing the
resistivity by 4 and 5 or
ders of maagnitude, respectively, respect to the
former cases.\\

Now, in order to consider simultaneously the band--gap energy and the
resistivity, we define

\begin{equation}
\xi(T) = \frac{\Delta\rho (T)}{\Delta Eg(T)} \ \ = \ \ \frac{\rho (80^oC)-
\rho(T)}{Eg(80^oC)-Eg(T)} \quad .
\end{equation}\\

The above parameter contains the information we are interested in.
In effect, high band--gap energy $Eg(T)$ is equivalent to low $\Delta Eg(T)$,
and similarly, low resistivity $\rho(T)$ implies high $\Delta\rho(T)$. Thus,
the higher $\xi(T)$ value, the better combination of high $Eg(T)$ and low
$\rho(T)$. According with it, we have found a procedure to find out the TTA
which produces samples with most favorable characteristics. We should observe
that $\xi(T)$ is basically governed by $\Delta\rho(T)$ because of its high
values compared to the $\Delta Eg(T)$ ones. It gives us an idea of what TTA
should be in advance investigated, namely, those with high $\Delta\rho(T)$
values.\\

We have calculated $\xi(T)$ for all the annealing atmospheres. With their
highest values, which are written in Table IV, we have  plotted the figure
of merit shown in Fig.4. This figure allows us to know that, among the
analysed annealings, the best effects are produced by H$_2$ and H$_2$+In
atmospheres. In particular, H$_2$+In annealing is lightly better than H$_2$
annealing, its best value $\xi= 4.6 \times 10^6 \Omega$ cm/eV taking place
at 250$^o$C. It is worth to point out that the measured resistivity of this
sample is $\rho= 5 \times 10^{-2} \Omega$ cm. A similar value has been
reported \cite{Toyotomi} but for the crystal grown by the melt, using the
Tamman method in Ar gas of about 110 atm pressure. Moreover, for the  Ar+S$_2$
atmosphere, the minimum resistivity is $\rho= 1.9 \times 10^3 \Omega$ cm.
This resistivity is roughly reproduced $(2.3 \times 10^3 \Omega$ cm) for a
normal CdS crystal but at 4.2$^oK$\cite{Toyotomi}.\\

The X-Ray diffraction patterns show that all the annealed samples present a
light interplanar--distance increase compared to the as--grown samples, but
only the samples exposed to Ar and Ar+S$_2$ annealings
undergo a cubic to hexagonal phase transition.  On the contrary, the samples
annealed in H$_2$, H$_2$+In and air atmospheres preserve the same structural
phase along the complete thermal process \cite{Melo}. The Mechanism that
explains this microestructural behavior is under study at present.\\

Photoluminiscence spectra of CdS samples annealed in Ar and Ar+S$_2$
atmospheres present a
shift of the luminescent peaks as a function of TTA. This shift has been
interpreted as an evidence of evolution from S$_2$ vacancies to interstitial
S$_2$ formation. For these two annealings, where a cubic to hexagonal
phase transition is completed, interstitial S$_2$ becomes part of the
hexagonal structure itself. PL spectra of CBD/CdS samples annealed in H$_2$,
H$_2$+In, and air atmospheres has not been performed up to now.\\

In summary we have found that CBD/CdS films present optimum conditions of
low resistivity and high band--gap energy when exposed to H$_2$+In annealing
atmosphere at 250$^o$C. This result allows us to know a possible procedure to
obtain low resistivity CdS/CdTe solar cells with high quantum efficiency,
therefore a real candidate for high conversion-efficiency solar cells.\\

\noindent{\LARGE Acknowledgements}\\

This work was partially supported by the Mexican agency CONACYT. H. V. and A.
F. S. acknowledge the Brazilian agency CNPq/MCT for financial support. A. F. S.
acknowledges CINVESTAV-IPN for its hospitality.
\pagebreak

\pagebreak

\noindent {\LARGE TABLE I}. Values of the band--gap energy, $E_g$, and
band--gap shift, $\Delta E_g$, as a function of temperature of thermal
annealing for Ar and Ar+S$_2$ annealings.

\vspace{3cm}
\begin{center}

\begin{tabular}{|c|c|c|c|c|}\hline
 \multicolumn{1}{|c}{Temperature ($^o$C)}
&\multicolumn{2}{|c|}{Ar}&\multicolumn{2}{|c|}{Ar+S$_2$}\\ \cline{2-5}
  &$Eg$(eV)&$\Delta Eg$(eV)&$Eg$(eV)&$\Delta Eg$(eV)\\ \hline
80 (as--grown)&2.42&--&2.42&-- \\ \hline
208&2.40&0.02&2.40&0.02\\ \hline
260&2.36&0.06&2.37&0.05\\ \hline
296&2.35&0.07&2.29&0.13\\ \hline
353&2.30&0.12&2.28&0.14\\ \hline
398&2.29&0.13&2.33&0.09\\ \hline
418&    &    &2.34&0.08\\ \hline
447&2.31&0.11&    &    \\ \hline
\end{tabular}
\end{center}
\newpage

\noindent{\LARGE TABLE II.} Values of the band--gap energy, $E_g$, and
band--gap
shift, $\Delta E_g$, as a function of temperature of thermal annealing for
H$_2$,
H$_2$+In and air annealings.

\vspace{3cm}
\begin{center}
\begin{tabular}{|c|c|c|c|c|c|c|}\hline
\multicolumn{1}{|c|}{Temperature ($^o$C)}&\multicolumn{2}{|c|}{H$_2$}
&\multicolumn{2}{|c|}{H$_2$+In}&\multicolumn{2}{|c|}{air}\\ \cline{2-7}
&$Eg$(eV)& $\Delta Eg$(eV)& $Eg$(eV)& $\Delta Eg$(eV)& $Eg$(eV)& $\Delta
Eg$(eV)\\
\hline
80(as--grown)&2.42&--  &2.42&--  &2.42&--\\ \hline
200&2.40&0.02&2.40&0.02&2.37&0.05\\ \hline
250&2.39&0.03&2.37&0.05&2.34&0.08\\ \hline
300&2.37&0.05&2.34&0.08&2.27&0.15\\ \hline
350&2.31&0.11&2.28&0.14&2.25&0.17\\ \hline
400&2.27&0.15&2.27&0.15&2.26&0.16\\ \hline
450&2.27&0.15&2.28&0.14&2.26&0.16\\ \hline
\end{tabular}
\end{center}
\newpage

\noindent {\LARGE TABLE III.} Resistivity minimum values and corresponding
$E_g$ as a function of temperature in different annealing atmospheres.

\vspace{3cm}
\begin{center}
\begin{tabular}{|c|c|c|c|}\hline
annealing & minimum $\rho$ & $E_g$ & temperature\\
          & ($\Omega$ cm) & (eV)   & ($^o$C)    \\ \hline
Ar        &2.5$\times 10^2$&  2.30 & 353        \\ \hline
Ar+S$_2$  &1.9$\times 10^3$&  2.29 & 296        \\ \hline
H$_2$     & 0.15           &  2.37 & 300        \\ \hline
H$_2$+In  & 0.05           &  2.37 & 250        \\ \hline
air       &1.0$\times 10^3$&  2.27 & 300        \\ \hline
\end{tabular}
\end{center}
\pagebreak

\noindent {\LARGE TABLE IV.} $\xi$(T) highest values in different annealing
atmospheres.

\vspace{3cm}
\begin{center}
\begin{tabular}{|c|c|c|}\hline
annealing&highest $\xi$&temperature\\
         &(10$^8 \Omega$ cm/eV)&($^o$C)\\ \hline
air&1.53&300\\ \hline
Ar+S$_2$&1.76&296\\ \hline
Ar&2.09&447\\ \hline
H$_2$&4.59&300\\ \hline
H$_2$+In&4.60&250\\ \hline
\end{tabular}
\end{center}

\pagebreak
\noindent {\LARGE Figure Captions.}\\

\begin{enumerate}
\item PA Normal Intensity of CdS films annealed in Ar+S$_2$ atmosphere for
different TTA as a function of energy.
\item Band--gap energy dependence on TTA for CdS annealed in Ar+S$_2$
atmosphere.
\item Band--gap shift dependence on TTA for CdS annealed in different
atmospheres.
\item $\xi(T)$ highest value for each different annealing atmosphere.
\end{enumerate}

\end{document}